\documentstyle[twoside,fleqn,npb,epsfig]{article}
\newcommand{\AmS}{{\protect\the\textfont2
  A\kern-.1667em\lower.5ex\hbox{M}\kern-.125emS}}

\title{Analytic evaluation of Feynman graph 
       integrals \thanks{ presented by E.R. at {\it RADCOR 2002 and Loops 
       and Legs in Quantum Field Theory}, 8-13 September 2002, Kloster Banz,
       Germany} } 

\author{P. Mastrolia 
        \address{Dipartimento di Fisica, Universit\`a di Bologna, 
                 I-40126 Bologna, Italy}
        \address{Institut f\"ur Theoretische Teilchenphysik, 
                 Universit\"at Karlsruhe, 
                 D-76128 Karlsruhe, Germany} 
        \thanks{e-mail mastrolia@bo.infn.it} 
        and 
        E. Remiddi$^{\ \rm a}$ 
        \address{INFN, Sezione di Bologna, I-40126 Bologna, Italy} 
        \thanks{e-mail remiddi@bo.infn.it} 
       } 

\begin{document}

\begin{abstract}
We review the main steps of the differential equation approach to 
the analytic evaluation of Feynman graphs, showing at the same time 
its application to the 3-loop sunrise graph in a particular kinematical 
configuration. \\ 
\end{abstract}

\maketitle

\noindent $\bullet$ 
The differential equation approach to the analytic evaluation of Feynman 
graph integrals applies to loop integrals defined in the by now customary 
regularization in $d$ continuous dimensions. We will recall the main 
steps of the approach, describing as an example its application 
to the 3-loop sunrise graph in the kinematical configuration of Fig.1. 
\includegraphics*[40mm,15mm][125mm,55mm]{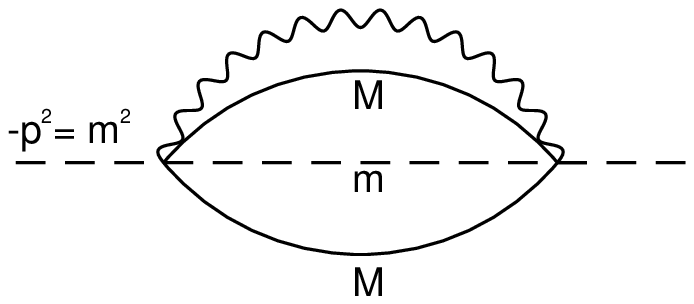} 
\centerline{ {Fig. 1. The considered 3-loop sunrise graph.} } 
\vskip 2mm 
\noindent 
$\bullet$ As the very first step, for any given graph one has to 
take all the scalar products which can be formed with the external 
momenta and all the loop momenta. In the case of Fig.1 there is a single 
external momentum $p$ and three loop momenta $k_i$, $i=1,2,3$, so that, 
irrespectively of the actual structure of the graph, the scalar 
products are the three $(p\cdot k_i)$, the three $k_i^2$ and the 
three $(k_i\cdot k_j)$ for $i\neq j$, for a total of nine. 
One can choose to express some of the scalar products as linear 
combination of the denominators (in general, there are several possible 
choices, equivalent for the following); the surviving scalar products 
are said irreducible (perhaps better not trivially reducible). One 
then considers the whole family of all the integrals obtained by taking 
as integrand all the possible combinations of the $P$ denominators of 
the scalar propagators raised to integer powers $\alpha_q$, $q=1,..,P$ 
and of the $S$ irreducible scalar products raised to integer 
powers $\beta_r$, $r=1,...,S$, the continuous dimensional regularization 
ensuring the absence of convergence problems. The corresponding 
integrals are functions of the continuous dimension $d$, of the Mandelstam 
variable $p^2$ and 
of the set of $(P+S)$ indices $\alpha_q, \beta_r$. (For Fig. 1, 
$P=4$ and $S=5$). For the following discussion, let us further define 
$A = \sum_q (\alpha_q-1)$, with $\alpha_q \ge 1$ and 
$B = \sum_r\beta_r$, with $\beta_r \ge 0 $. \\ 
$\bullet$ As next step, we look for relations between the previous 
integrals. The relations are best provided by means of the by now famous 
integration by parts identities (\emph{ibp-id's})
~\cite{IBP} , obtained by replacing in any 
of the above integrals the integrand by all the combinations of the 
kind $ (\partial/\partial k_{i,\mu})v_\mu $, times the 
same integrand, where $i=1,..,3$ and $v_\mu$ is any of the external or 
loop momenta. 
(For the graph of Fig, 1 there are $3\times 4 = 12$ different 
combinations $ (\partial/\partial k_{i,\mu})v_\mu $.) \\ 
By carrying out explicitly 
the derivatives, and after some trivial algebra, one is left with 
a combination of polynomials (depending in general on $d$, on the 
Mandelstam variables, and on the internal masses, always with integer 
coefficients) times integrals of the same family. But the whole 
original integral vanishes, as the integrand is by construction 
a divergence, so that the combination of integrals vanishes as well 
-- an integration by parts identity has been obtained. 
If the original integrand belongs to the class $(A,B)$, {\it i.e.\ } 
if the sums of the powers of denominators and numerators are $A+P$ and
$B$ respectively, 
the involved integrals belong to the classes $(A\pm1,B\pm1)$. 
Integrals in which one of the propagator is missing ({\it i.e.\ } its 
exponent is 0) are said to correspond to subtopologies, and can be 
considered as already known in a systematic bottom up approach. 
(It is to be observed that other identities of similar nature can be 
obtained by exploiting the 
Lorentz Invariance ~\cite{LI} on the external vectors, but only when there 
are at least three independent external vectors.) \\ 
$\bullet$ At this point, it is natural to try to exploit the identities 
for expressing as many as possible of the integrals actually needed in any 
explicit calculation in terms of as few as possible integrals of the set 
-- the so-called master integrals. 
The original ideas for building algorithms applying to the most general 
case goes back to Laporta (\cite{LI}-\cite{Bastei}): 
the integrals are ordered by giving them a 
\emph{weight} (that can be done almost at will by giving a higher weight 
to the integrals considered more complicated; typically, the fully 
scalar integral is given the lowest weight), and then a ``sufficient 
number" of explicit equations is written for them. 
In the case of Fig. 1, starting from the fully scalar integral, 
corresponding to $(A=0,B=0)$, one obtains 12 \emph{ibp-id's} involving, 
besides the original integral, the 4+5+20 = 29 integrals with $(A=1,B=0)$, 
$(A=0, B=1)$ and $(A=1, B=1)$ -- an apparently runaway situation, in 
which the involved integrals outnumber the equations. But that is 
not the case; an elementary combinatorial calculation shows that 
by writing all the \emph{ibp-id's} starting from $A=0,1$ and 
$B=0,1$ one generates $12\times 30 = 360$ equations, involving 
integrals with $A=0,1,2$ and $B=0,1,2$ at most; as the number of all 
the integrals with $A,B=0,1,2$ is just 336 
one has more equations than integrals, {\it i.e.\ } an apparently 
overconstrained system. That is not the case either - simply, the equations 
are not all linearly independent. \\ 
The system can be solved by Gauss substitution, eliminating first the 
integrals of higher weight -- the method is 
elementary, even if, given the size of the system, it is convenient to 
use an automatized computer procedure. When that is done, it is found 
that all the integrals appearing when dealing with the graph of 
Fig.1. can be expressed in terms of 4 Master Integrals (MI's). \\ 
$\bullet$ Let $F(d,p^2)$ be any of the MI's. Obviously, 
$$ p^2\frac{\partial}{\partial p^2} F(d,p^2) = 
   2 p_\mu\frac{\partial}{\partial p_\mu} F(d,p^2) . $$ 
Substitute in the {\it r.h.s.} for $F(d,p^2)$ its original definition 
as integral on the loop momenta of a suitable integrand. By carrying out 
explicitly the derivative $\partial/\partial p_\mu$ and then contracting 
with $p_\mu$ one obtains a combination of integrals associated to the 
considered graphs, which can then be expressed, thanks to the solution 
of the \emph{ibp-id's} discussed in the previous paragraph, as combination 
of the MI's. From the above equation one has therefore expressed 
the $p^2$-derivative of the particular MI $ F(d,p^2) $ in terms of 
the MI's of the problem. 
By repeating the procedure for all the MI's, one obtains 
a system of linear first order differential equations for the MI's. 
The system is in general non homogeneous, as the \emph{ibp-id's} can 
introduce integrals where some of the propagators are missing. 
The non-homogenous part involves simpler graphs, and can therefore be 
considered as known in a systematic bottom up approach. \\ 
If the MI's depend on several independent momenta $p^i_\mu$ and 
correspondingly by several Mandelstam variables, by acting on 
the MI's with all the combinations $p^i_\mu \partial/\partial p^j_\mu$ and 
minor algebraic rearrangements one can obtain a system of differential 
equations in any of the Mandelstam variables ~\cite{DiffEq}. 
It is to be recalled here that the solutions of the \emph{ibp-id's} can 
be almost immediately rewritten as differential equations in the 
internal masses ~\cite{Kotikov}. \\ 
Finally, a linear system of differential equations can always be rewritten 
as a single differential equation of suitable higher order for any of its 
unknow functions. 
The coefficients appearing in the equations are in any case 
polynomials (or rational factors, when dividing the equations by the 
polynomial multiplying the highest derivative). \\ 
$\bullet$ In the case of the special kinematical 
situation depicted in Fig.1, one of the internal 
masses vanishes, two internal masses take the same value $M$, 
the fourth the value $m$ and the Mandelstam variable is timelike and 
takes the value $m^2$, {\it i.e.\ } $p^2=-m^2$. We rescale all the 
momenta by $M$, further put $ m = Mx $ and define 
{\small
\begin{eqnarray} 
 \Phi(d,x) = \frac{ C^{-3}(d) }{ (2 \pi)^{3(d-2)} } 
     \int 
      d^dk_1 \ d^dk_2 \ d^dk_3 
     \ \times & & \nonumber \\ 
         \frac{1}{ k_1^2 
              (k_2^2+1)
              (k_3^2+1)
              [(p-k_1-k_2-k_3)^2+x^2]}, && 
\label{eq:defF} 
\end{eqnarray} 
} %% end-small 

\noindent
where $C(d) = (4\pi)^{2-d/2}\ \Gamma(3-d/2) $ is a normalization 
factor, introduced for convenience, with the limiting value 
$C(4)=1$ at $d=4$. 
The above discussed equations in $p^2$ and the masses give for 
$\Phi(d,x)$ the following 3rd O.D.E., exact in $d$: 
\begin{eqnarray} 
%%  \sum_{j=1}^3 P_j(d,x) \frac{d^{ j }}{dx^j} \Phi(d,x) &  & \nonumber \\ 
%% + P_0(d,x) \Phi(d,x) & = & N(d,x) \ ,
   \Big[   \sum_{j=1}^3 {\mathcal{P}}_j(d,x) \frac{d^{ j }}{dx^j} 
       + {\mathcal{P}}_0(d,x) \Big]  \Phi(d,x) = && \nonumber \\ 
   {\mathcal{N}}(d,x), && 
\label{eq:big} 
\end{eqnarray} 
where
\begin{eqnarray}
{\mathcal{P}}_3(d,x) & = &  x^2 (1-x) (1+x) \ ;  \nonumber \\
{\mathcal{P}}_2(d,x) & = &   \big[ 2 - (d-4)\big] x 
                + \big[ 2 + 5 (d-4) \big] x^3    \ ;   \nonumber \\
{\mathcal{P}}_1(d,x) & = & - \big[ 6 + 8(d-4) + 2(d-4)^2 \big]  \nonumber \\
         &   & + \big[ 2 - 4(d-4) - 6(d-4)^2 \big] x^2 \ ;   \nonumber \\
{\mathcal{P}}_0(d,x) & = & 
                  - \big[ 8 + 14(d-4) + 6(d-4)^2 \big] x  \ ;  \nonumber \\
{\mathcal{N}}(d,x)   & = & 
                    \frac{1}{(d-4)^3} \frac{x^{(d-2)}}{x}   \ . 
\end{eqnarray}
$\bullet$ The direct solution of the differential equation, for 
arbirary $d$, seems out of reach. But as, in any case, one is interested 
in the $d\to 4$ limit of the solution, one can expand $ \Phi(d,x) $ 
in Laurent series in $(d-4)$ and rewrite the equation in terms of 
the coefficients of the expansion. Let us take as known, for simplicity, 
that in our case the leading (most singular) term is of order 
$1/(d-4)^3$, so that the expansion of $ \Phi(d,x) $ reads  
\begin{equation} 
  \Phi(d,x) = \sum_{k \ge -3} (d-4)^k \Phi^{(k)}(x) \ . 
\label{eq:expa} 
\end{equation} 
By correspondingly expanding Eq.(\ref{eq:big}) in $(d-4)$, one obtains 
a system of chained equations for the $ \Phi^{(k)}(x) $; the first 
equation involves only $ \Phi^{(-3)}(x) $, and is to be solved for 
$ \Phi^{(-3)}(x) $ independently of the other coefficients; the second 
equation is an equation for $ \Phi^{(-2)}(x) $ involving 
$ \Phi^{(-3)}(x) $ in the non-homogeneous term, and so on. 
The general equation reads 
\begin{eqnarray} 
 & & \Big[ \sum_{j=1}^3 P_j(x)\frac{d^{ j }}{dx^j} 
            + P_0(x) \Big] \Phi^{(k)}(x) \nonumber \\
 &=& \Big[ \sum_{l=1}^2 Q_l(x)\frac{d^{ l }}{dx^l} 
            + Q_0(x) \Big] \Phi^{(k-1)}(x) \nonumber \\
 & & + \Big[  R_1(x)\frac{d}{dx} 
            + R_0(x) \Big] \Phi^{(k-2)}(x) \nonumber \\
 & & + S_0(x) \frac{1}{(k+3)!} \ln^{(k+3)}(x)
\label{eq:bigger} 
\end{eqnarray} 
where $\Phi^{(k)}(x) = 0$ if $k<-3$, and $P_i(x)$, $Q_i(x)$, $R_i(x)$, 
$S_0(x) $ are simple polynomials in $x$ whose explicit expressions 
can be easily derived from Eq.(\ref{eq:big}). \\ 
All the above equations have the same form 
\begin{equation} 
{\mathcal{D}}(x) \Phi^{(k)}(x) = N^{(k)}(x) \ , 
\label{eq:eqk} 
\end{equation} 
where ${\mathcal{D}}(x)$ is the differential operator applied to 
$\Phi^{(k)}(x)$ in the {\it l.h.s.} of Eq.(\ref{eq:bigger}) and 
$N^{(k)}(x)$ stands for the non-homogeneous part, to be considered 
as known when dealing with the order $k$ in the $(d-4)$ expansion. 
The homogeneous part, which is the same for any $k$, can be written as 
\begin{eqnarray} 
 & &  {\mathcal{D}}(x) \phi(x) \ \equiv \nonumber \\ 
 & &  \left[   \frac{d^{ 3 }}{d x^3} 
         + 2 \ \left(   \frac{1}{x} 
                      + \frac{1}{1-x} 
                      - \frac{1}{1+x} \right) 
                    \frac{d^{ 2 }}{d x^2} \right. \nonumber \\ 
 & &   \ - 2 \ \left(    \frac{3}{x^2} 
                      + \frac{1}{1-x} 
                      + \frac{1}{1+x} \right) 
                    \frac{d}{d x} \nonumber \\ 
 & &  \left.
      \ - 2 \ \left(   \frac{3}{x^2} 
                      + \frac{1}{1-x} 
                      + \frac{1}{1+x} \right) 
                    \right] \phi(x) \nonumber \\
 & &  = 0 \ .
\label{eq:homo} 
\end{eqnarray} 
$\bullet$ We will try to solve the equations for the coefficients of the 
$(d-4)$ expansion by means of the method of the variation of constants 
by Euler. To that aim, we need the solutions of the homogeneous equation, 
Eq.(\ref{eq:homo}). \\ 
Up to here, our approach applies to virtually any Feynman graph integral; 
for continuing one has to solve the homogeneous differential equation 
in $d=4$ dimensions. Algorithms for the solution of differential equations 
in the general case are unfortunately not available. But as a more 
optimistic remark, let us point out that in almost all the cases 
considered sofar (see for instance ~\cite{(non)planar}, 
~\cite{sun2},~\cite{sun3}) the homogeneous equations came out to be 
rather simple, with solutions which could be expressed in terms of 
almost elementary functions. The same happens also in the present case. 
Indeed, it is immediately seen that a first solution of Eq.(\ref{eq:homo}) 
is 
$$ \phi_1(x) = (1 - x^2) \ ; $$ 
by putting $ \phi(x) = \phi_1(x) \psi(x) $ one finds a second order 
equation for $\psi'(x) $ having the simple solution 
$$ \psi'_1(x) = (1-x^2)^2/x^3 \ , $$ 
and finally, putting  $ \psi'(x) = \psi'_1(x) \chi(x) $, one finds 
at once 
$$ \chi'(x) = - x^4/(1 - x^2)^5 \ .$$ 
The corresponding solutions are 
\begin{eqnarray}
\phi_2(x) & = & \nonumber \\
   &   & \hspace*{-1.5cm}
          - \ \frac{1}{2} \frac{(1 - x^2) (1 - x^4) }{ x^2 } 
          - 2 (1 - x^2) H(0;x)  \ , \nonumber \\
 &  &  \nonumber \\ 
\phi_3(x) & = & \nonumber \\
   &   & \hspace*{-1.5cm}
          + \ \frac{3}{512} \frac{(1 - x^2) (1 - x^4) }{ x^2 }
                  \ [H(-1;x) + H(1;x)] \nonumber \\ 
   &   & \hspace*{-1.5cm} 
          + \frac{3}{128} (1 - x^2) \ [H(0,-1;x) + H(0,1;x)]  
                \nonumber \\ 
   &   & \hspace*{-1.5cm}
                - \ \frac{1}{256} \frac{(x^2 + 3) (3 x^2 + 1)}{x} \ ,
         \nonumber 
\end{eqnarray}
where the functions $H$ are Harmonic Polylogarithms or HPLs~\cite{RemVer}. 
Referring to ~\cite{RemVer} for more details, let us recall that the 
HPLs depend on an argument, say $x$, and on a vector of indices, say 
$\vec{b}$, whose components take any of the values $(1,0,-1)$, and 
whose number is called the weight ($w$). At $w=1$, they are defined as 
\begin{eqnarray} 
    H(0;x) &=& \ln(x) \ , \nonumber\\ 
    H(1;x) &=& -\ln(1-x) = \int_0^x \frac{dx'}{1-x'} \ , \nonumber\\ 
    H(-1;x) &=& \ln(1+x) = \int_0^x \frac{dx'}{1+x'} \ ; \nonumber 
\end{eqnarray} 
at higher weight, $w>1$, if the vector of the indices is written as 
$(a,\vec{b})$, where $\vec{b}$ is a vector of $w-1$ components, 
they fulfill 
\begin{equation} 
\frac{d}{dx} H(a,\vec{b};x) = f(a,x)H(\vec{b};x) \ , 
\label{def:HPL} 
\end{equation} 
with 
\begin{eqnarray} 
    f(1,x) &=& 1/(1-x) \ , \nonumber\\ 
    f(0,x) &=& 1/x \ , \nonumber\\ 
    f(-1,x) &=& 1/(1+x) \ . 
\label{eq:deff} 
\end{eqnarray} 
\noindent 
The Wronskian of the $\phi_i(x)$, 
as expected by direct inspection of Eq.(\ref{eq:homo}), is 
$$ W(x) = - (1 - x)^2 (1 + x)^2/x^2 \ . $$ 
In terms of the solutions of the homogeneous equation, the 
general solution of the inhomogeneous equation is given by 
\begin{eqnarray} 
  \Phi^{(k)}(x) &=& \nonumber \\ 
 & & \hspace*{-1.85cm} 
  \phi_a(x) 
         \Big[ \Phi^{(k)}_a  
         + \int^x \ dy \ 
           \frac{ \epsilon_{abc} \ M_{bc}(y)}{W(y)} N^{(k)}(y) \Big]
                    \ , \nonumber\\ 
 M_{bc}(y) &=& \phi_b(y)\phi_c^{'}(y) - \phi_b^{'}(y)\phi_c(y) \ ,
\label{eq:Euler} 
\end{eqnarray} 
where $\{a,b,c\}$ is a permutation of $\{1,2,3\}$ and $\Phi^{(k)}_a (a=1,2,3)$
are three integration constants to be fixed, order by order in $k$, 
to match the boundary conditions identifying $\Phi(d,x)$ 
Eq.(~\ref{eq:defF}). \\ 
$\bullet$ Let us look at Eq.(\ref{eq:Euler}) for the first value of $k$, 
which is $k=-3$. From Eqs.  (\ref{eq:bigger}) one gets 
$$ N^{(-3)}(x) = \frac{1}{2} \left( \frac{2}{x} + \frac{1}{1-x} 
               - \frac{1}{1+x} \right)  \ ; $$ 
when it is inserted in Eq.(\ref{eq:Euler}), 
% one is left with a combination of 
% powers of the monomials $x, (1-x)$ and $(1+x)$. 
after partial fractioning 
and some trivial integration by parts, the remaining non trivial 
integrals involve only one of the three factors $1/x, 1/(1-x), 1/(1+x) $, 
the three factors $f(a,x)$ of Eq.(\ref{eq:deff}) 
and polylogarithms of various weight: but those (indefinite) integrals 
can be carried out at once by using Eq.(\ref{def:HPL}).
It follows that $ N^{(-2)}(x), $ which contains $ \Phi^{(-3)}(x), $ 
is also a combination of powers of the monomials $x, (1-x), (1+x)$ and 
polylogarithms, so that the procedure can be iterated at will up to 
any desired value of $k$, the result being, at all orders, a combination 
of harmonic polylogarithms. \\ 
$\bullet$ To complete the job, we must fix the sofar arbitrary constants 
$ \Phi^{(k)}_a, a=1,2,3. $ To that aim, let us observe that the behaviour 
of the most general solution of Eq.(\ref{eq:big}) is of the kind 
$$ \Phi(d,x) = \sum_{i=1}^4 C_i x^{\alpha_i}\left( 1 + c_{i,1}x^2 
             + c_{i,2}x^4 + .... \right) \ , $$ 
where the four possible values of the exponents are 
$\alpha_1=0 $, $\alpha_2=-(d-2) $, $\alpha_3=(2d-5) $, 
and $\alpha_4=(d-2) $, as can be easily verified by substituting 
the above expansion in Eq.(\ref{eq:big}). On the other hand, by direct 
inspection of $\Phi(d,x)$ Eq.(\ref{eq:defF}), one finds that the 
behaviours $\alpha_2$ and $\alpha_3$ are ruled out, as the integral does not 
diverge in $x$ in the $x\to 0$ limit when $d>2$; this implies that 
terms like $1/x^2$ or $1/x$ (or more in general terms with odd powers 
of $x$) cannot be present in the expansion around $x=0$ in the 
$d\to4$ limit. Similarly, one finds by inspection that $\Phi(d,x)$ 
is analytic in $x$ for $x\to 1$; this implies that terms in $\ln(1-x)$ 
cannot be present in the expansion around $x=1$ in the $d\to4$ limit. 
Those conditions are sufficient to fully determine the integration 
constants. Indeed, we evaluated explicitly Eq.(\ref{eq:Euler}) up to $k=5$, 
and by imposing the proper behaviours for $x\to 0$, $x\to 1$ we 
fixed the constants
$\Phi^{(k)}_1$, $\Phi^{(k)}_2$, $\Phi^{(k)}_3$ for $-3\le k \le 3$, 
($\Phi^{(k)}_3$ being 0 for any $k$) and $\Phi^{(4)}_2$. 
The constants $\Phi^{(4)}_1$, $\Phi^{(5)}_1$ and
$\Phi^{(5)}_2$ remained still undetermined - to fix them, 
one has to impose the boundary conditions to further terms 
$\Phi^{(k)}(x)$ with $k = 6,7$. \\ 
The explicit analytic solution up to $k=3$ included, which we obtained 
in that way, involves HPL's of argument $x$ and up to weight $w=6$. 
The resulting expression is unfortunately too long to be listed here. 
More details can be found in~\cite{sun3}. 
\\ 
$\bullet$ Having the full dependence on $x$ in closed analytic form it is 
immediate to obtain the values at $x=0$ and $x=1$. The $x=0$ values 
can also be obtained by direct integration of Eq.(\ref{eq:defF}); they 
can be used as a check of the calculation. \\ 
The values at $x=1$ are more interesting. They correspond to the 
quantity $J_{11}$ of~\cite{3lgm2}, where they were given up to order 
$(d-4)^3$; from our solution (and from the table~\cite{VermTab} of the 
values of the HPL's at $x=1$), we could evaluate them up to order 
$(d-4)^5$ included; in fact, given the structure of the solutions 
of the homogeneous equation, the still undetermined constants are 
not needed for the $x=1$ value. 
The results up to $(d-4)^3$ are in full agreement with ~\cite{3lgm2}; 
the new terms, namely the coefficients of order 
$(d-4)^4$ and $(d-4)^5$ match the numerical values of the 
quantity $I_{12}$ of ~\cite{Lap3l}. 
\\ \\ 
{\bf Acknowlegments.} \\ We wish to thank J. Vermaseren for his kind 
assistance in the use of his algebraic program  
{\tt FORM}~\cite{FORM}, by which all the calculations were carried out. 
%%%%%%%%%%%%%%%%%%%%%%%%%%%%%%%%%%%%%%%%%%%%%%%%%%%%%%%%%%%%%%%%%%%%%%%

%%%%%%%%%%%%%%%%%%%%%%%%%%%%%%%%%%%%%%%%%%%%%%%%%%%%%%%%%%%%%%%%%%%%%%%
\end{document}